\documentclass[a4paper,fleqn,usenatbib]{mnras}
\usepackage{graphicx}
\usepackage{subfigure}
\usepackage{amsmath}
\usepackage{amssymb}
\usepackage{sidecap}
\usepackage{color}
\usepackage{enumerate}
\usepackage{enumitem} 
\usepackage{algorithm}
\usepackage{algpseudocode}
\usepackage{verbatim}

\def\ba{\begin{eqnarray}}
\def\ea{\end{eqnarray}}
\def\be{\begin{equation}}
\def\ee{\end{equation}}

\providecommand{\adsurl}[1]{\href{#1}{ADS}}

\newcommand\Msun{$ {\rm M}_{\odot}$} 

\title[Ultra-Light Dark Matter in Ultra-Faint Dwarf Galaxies]{Ultra-Light Dark Matter in Ultra-Faint Dwarf Galaxies}
\author[Erminia Calabrese and David N. Spergel]{Erminia~Calabrese$^{1,2}$\thanks{erminiac@astro.princeton.edu} and David~N.~Spergel$^{1}$\thanks{dns@astro.princeton.edu}\\
$^{1}$ Department of Astrophysical Sciences, Peyton Hall, Princeton University, 4 Ivy Lane, Princeton, NJ USA 08544\\
$^{2}$ Sub-department of Astrophysics, University of Oxford, Denys Wilkinson Building, Oxford, OX1 3RH, UK}

\begin{document}

\maketitle 

\begin{abstract}
Cold Dark Matter (CDM) models struggle to match the observations at galactic scales. The tension can be reduced either by dramatic baryonic feedback effects or by modifying the particle physics of CDM. Here, we consider an ultra-light scalar field DM particle manifesting a wave nature below a DM particle mass-dependent Jeans scale. For DM mass $m\sim10^{-22}{\rm eV}$, this scenario delays galaxy formation and avoids cusps in the center of the dark matter haloes. We use new measurements of half-light mass in ultra-faint dwarf galaxies Draco II and Triangulum II to estimate the mass of the DM particle in this model. We find that if the stellar populations are within the core of the density profile then the data are in agreement with a wave dark matter model having a DM particle with $m\sim 3.7-5.6\times 10^{-22}{\rm eV}$. The presence of this extremely light particle will contribute to the formation of a central solitonic core replacing the cusp of a Navarro-Frenk-White profile and bringing predictions closer to observations of cored central density in dwarf galaxies. 
\end{abstract}


\section{Introduction}
\label{sec:intro}
The $\Lambda$CDM model emerged in the last two decades as the simplest model that consistently accounts for most astrophysical and cosmological observations (\citealp{Spergel2003}, \citealp{PlanckP2015}). In this scenario most of the matter content of the Universe is in the form of a non-interacting and non-relativistic matter component, Cold Dark Matter (CDM), and at present its nature is unknown. 

CDM models successfully reproduce the evolution of a smooth early Universe into the cosmic structures observed today on a wide range of redshifts and scales. However, the agreement between models and observations degrades rapidly when zooming into the innermost galactic regions. 

In the standard model of galaxy formation, galaxies are seeded by dark matter haloes \citep{White78,Blum84,White91}, whose structure is a very powerful probe to distinguish between different theoretical models. DM haloes have been extensively studied with N-body simulations (see e.g, \citealp{Kuhlen2012} for a review) and the similarities between different CDM-only simulated haloes have justified the definition of a universal dark matter halo profile, the Navarro-Frenk-White (hereafter NFW, \citep{NFW}) profile. However, improved resolution in N-body simulations has revealed that at small scales the CDM paradigm presents three major problems (see \citealp{Weinberg2013} and references therein): (i) the cusp-core problem - NFW density profiles arising from CDM-only simulations predict a steeper density (cusp) towards the center of the DM halo compared to disc and dwarf spheroidal galaxies observations of a flatter central density (core); (ii) the missing satellites problem - CDM models predict more-than-observed Milky Way satellite galaxies living in DM sub-haloes; (iii) the too-big-too-fail problem - CDM models predict more-than-observed massive DM sub-haloes.

These problems can be solved at the theory level with two possible approaches: including baryonic feedback or other astrophysical effects into the simulations (see e.g., \citealp{Governato2012,DiCintio2014,Pontzen2014,Pawlowski2015,Orbe2015,Papastergis2015} for recent discussions), or modifying the CDM component. 

Here, we take the latter approach and consider a modification of the particle physics of DM, allowing for the presence of a light bosonic, axion-like, dark matter particle. If the DM particle is ultra-light, with mass $\sim10^{-22} {\rm eV}$, than its wave nature can manifest on astrophysical scales and bring theoretical predictions closer to the observations \citep{Goodman2000,Hu00,Schive1,Schive2,Marsh2013}. 

Ultra-light axion-like particles are one of the most compelling candidates for CDM and have been explored with different observables covering many cosmic epochs (from Cosmic Microwave Background data \citep{Hlozek2014}, Lyman-$\alpha$ systems \citep{Amendola2006}, reionization history \citep{Bozek2014,Schive3,Sarkar2015}, galaxy formation and dwarf galaxy dynamics \citep{Schive1,Lora2014,Lora2015,MPop2015}). Axions will behave like matter in the present Universe as long as their mass is $>10^{-33}{\rm eV}$ and consistency with the CMB demands that the dominant DM component has $m>10^{-24}{\rm eV}$ \citep{Hlozek2014}. On the other end, axions can feasibly be distinguished from CDM as long as $m<10^{-18}{\rm eV}$ \citep{Marsh2}; heavier axions are allowed as DM, but are indistinguishable form CDM in their effects on structure formation. Collecting all cosmological and astrophysical information we then currently know that $10^{-24}<m<10^{-18}{\rm eV}$.
See \citealp{Marsh} for a comprehensive review. 

To add to these DM mass constraints here we consider two recently discovered ultra-faint dwarf galaxies in the Milky Way, Draco II and Triangulum II \citep{Laevens20150,Laevens2015}. Both objects seem to be very peculiar. The analysis of \citealp{Martin2015} classifies Draco II as the smallest dwarf galaxy ever confirmed. Triangulum II is bigger than Draco II but fainter. Its nature is still controversial (see discussion in \citealp{Kirby2015} and \citealp{Martin2}) but preliminary data reductions suggest it is likely to be a dwarf galaxy. In particular, it seems to be the dwarf with the largest mass-to-light ratio and so the most DM dominated object ever observed. These two galaxies will complement existing constraints on DM.

The paper is organized as follows. We describe the wave dark matter model in Section~\ref{sec:wdm} and the data in Section~\ref{sec:ufdg}. In Section~\ref{sec:m22} we estimate the DM particle mass and compare the predictions with a $\Lambda$CDM NFW model. We discuss our results and conclude in Section~\ref{sec:disc}.

\section{Wave Dark Matter Haloes}
\label{sec:wdm}
An alternative to CDM that has recently gained much attention is the Bose-Einstein Condensate Scalar Field Dark Matter model (BEC/SFDM) (see e.g., \citealp{Turner1983,Ji1994,Lee1996,Guzman1999,Goodman2000,Guzman2000,Bohmer2007,Sikivie2009,Woo2009,Lundgren2010,Harko20112,Harko2011,Chavanis2011,Chavanis2,Chavanis2012,Magana2012,Harko2012,Harko20122,Lora2012,Suarez2013,Lora2014,Li2014,Harko2014,Lora2015,MPop2015,Guzman2015,Harko2015,Harko20152,Davidson2015,Guth2015} and references therein), also known as Fuzzy Dark Matter \citep{Hu00} or Wave Dark Matter \citep{Schive1}. In this scenario DM is made of extremely light bosons, such as axion-like particles, non-thermally generated and described by a scalar field $\psi$. If the field's bosons are ultra-light, with a mass $\sim10^{-22} {\rm eV}$, quantum pressure from the bosons occupying the same ground state counters gravity and in the early Universe they condensate in a single coherent macroscopic wave (a self-gravitating Bose-Einstein condensate). The field is minimally coupled to gravity and interacts only gravitationally with the baryonic matter. In the Newtonian approximation and in the case of negligible self-interaction between bosons (as in the case presented in \citealp{Hu00,Schive1} that we will follow here), the scalar field is then well described by the coupled Schr{\"o}dinger and Poisson equations (see e.g., \citealp{Widrow93}) and DM haloes are the ground-state solution -- the gravitational configuration in equilibrium -- of the system.

A very interesting feature of this model is an effective Jeans scale depending on the DM particle mass below which the uncertainty principle counters gravity. Galactic haloes form by the usual gravitational instability but with perturbations suppressed below a scale that is set by the particle mass. As a result, interestingly for galaxy formation, for $m\sim10^{-22} {\rm eV}$ flat (cored) density profiles are produced within $\sim0.1-1.0$ ${\rm kpc}$, galaxy formation is delayed and the halo mass function shows suppression of haloes smaller than $\sim 10^{10}$\Msun, helping with both the cusp-core and missing satellites problems \citep{Goodman2000,Hu00,Schive1,Schive3,Marsh2013}.

In this work we consider the wave dark matter model, $\psi$DM, presented in \citealp{Schive1,Schive2}. \citealp{Schive1} performed the first high-resolution cosmological simulations for wave dark matter and showed that the central density profiles of all collapsed objects are well described by the stable soliton solution of the Schr{\"o}dinger-Poisson equation. A gravitationally self-bound soliton core is found in every halo superposed on a NFW profile. The NFW behaviour is recovered at larger radii and the $\psi$DM cosmology is indistinguishable from CDM on large scales.

Fitting cosmological simulations \citealp{Schive1} shows that the density profile of the innermost central region at redshift $0$ is well approximated by:
\be
\rho_s(r)=\frac{1.9 (10 m_{22})^{-2}r_c^{-4}}{[1+9.1\times10^{-2}(r/r_c)^2]^8} 10^9 {\rm M}_{\odot} {\rm kpc^{-3}} \,,
\ee
where $m_{22}\equiv m/10^{-22}{\rm eV}$ is the DM particle mass and $r_c$ is the core radius of the halo. The soliton extends up to $\sim3r_c$ \citep{Schive1,Schive2}.

From the density profile we can estimate the enclosed mass at a given radius $r$\footnote{We note that for this density profile the enclosed mass integral cannot be solved analytically and we will perform numerical estimates.}:
\be
M( < r) = \int_0^r 4 \pi \rho_s(r') r'^2 dr' \,. 
\ee
$M_c \equiv M(<r_c)$ gives the central core mass and the total soliton mass will be $\sim 4 M_c$.

\citealp{Schive2} also derive an analytical dependence between the core mass or radius and the total mass of the halo, $M_h$, hosting the galaxy. At the present time these relations are:
\be
M_c\sim \frac{1}{4} M_h^{1/3}(4.4\times10^7 m_{22}^{-3/2})^{2/3} \,,
\label{eq:mc}
\ee
\be
r_c \sim 1.6 m_{22}^{-1} \Big (\frac{M_h}{10^9 {\rm M}_\odot}\Big)^{-1/3}  {\rm kpc} \, ,
\label{eq:rc}
\ee
so that, for a given $m_{22}$, the largest cores are embedded on the smallest haloes.

For our purposes the soliton is the only component of the density profile that compares to the data. We assume that the stellar populations are within the core and we do not need any extrapolation to NFW at larger scales\footnote{This approach follows the analysis of \citealp{Schive1,Schive2} that found the soliton to be a good approximation of the full density profile. \citealp{MPop2015} included the NFW component and found that it was unconstrained, and the data could be fit using only the soliton.}.

\section{Ultra-Faint Dwarf Galaxies}
\label{sec:ufdg}
Dwarf galaxies are believed to be the most common type of galaxies in the Universe. In a hierarchical formation scenario these objects are the building blocks of more massive galaxies and are believed to have been even more numerous at earlier times\footnote{This is not necessarily the case in the $\psi$DM model where structure formation is not strictly hierarchical because of the mass function cut-off.}. The large mass-to-light ratios of dwarf galaxies, and in particular of ultra-faint dwarfs (half-light radius $\lesssim 50{\rm pc}$), suggest that they are the most DM dominated objects and therefore a great laboratory to test DM models. 

The richest source of information are nearby dwarfs, e.g., the Milky Way dwarf satellites, where individual stars can be resolved and stellar dynamics can track the density profile and the gravitational field. A candidate object identified in a survey can be considered a dwarf galaxy, and distinguished from stellar clusters, if it shows a velocity dispersion in excess of what would be expected from stellar mass alone ($\sim 0.3$ ${\rm km\, s^{-1}}$) or a dispersion in stellar metallicity ($\gtrsim 0.1$ dex in iron) indicating chemical self-enrichment \citep{Willman2012}. These two properties make the candidate a dwarf galaxy dominated by dark matter.

In the last two decades many faint dwarf galaxies have been discovered with photometric surveys (SDSS \citep{Belokurov2007}, DES \citep{Des1,Des2,Koposov2015}). Here we consider newly discovered ultra-faint dwarf galaxies, Draco II and Triangulum II, with the Pan-STARRS1 3$\pi$ survey \citep{Laevens20150,Laevens2015}.

\begin{itemize}
\item{Draco II}\\
Draco II discovery \citep{Laevens2015} reported a close (distance from the Sun $20\pm3$ $\rm{kpc}$), extremely compact (half-light radius $r_{h}=19^{+8}_{-6} {\rm pc}$) and faint ($L_{\odot} = 10^{3.1\pm0.3}$) object whose nature was uncertain. The addition of spectroscopic observations led \citealp{Martin2015} to find a small, yet marginally detected, velocity dispersion, $\sigma_{vr} = 2.9\pm2.1$ ${\rm km\, s^{-1}}$, and a highly sub-solar metallicity, $[\rm{Fe/H}]<-2.1$.  Draco II stellar dynamics constrain the half-light mass to be $\log_{10}(M_{1/2})=5.5^{+0.4}_{-0.6}$ and a mass-to-light ratio $\log_{10}((M/L)_{1/2}) = 2.7^{+0.5}_{-0.8}$ (\citealp{Martin2015}, using the definition of $M_{1/2}$ presented in \citealp{Wolf2010}). These estimates hint to a strongly dark matter dominated system and in particular to the smallest compact dwarf galaxy ever confirmed. \\

\item{Triangulum II}\\
Similarly to Draco II, Triangulum II photometric properties were first reported in \citealp{Laevens20150} and then followed-up with a spectroscopic analysis in \citealp{Kirby2015} and \citealp{Martin2}. Triangulum II is a larger ($r_{h}=34^{+9}_{-8} {\rm pc}$) but fainter ($L_{\odot} = 10^{2.6\pm0.2}$) system located at $30\pm2$ $\rm{kpc}$. The velocity dispersion of the member stars was found to be $\sigma_{vr} = 5.1^{+4.0}_{-1.4}$ ${\rm km\, s^{-1}}$ with a corresponding $\log_{10}(M_{1/2})=5.9^{+0.4}_{-0.2}$ and $(M/L)_{1/2} = 3600^{+3500}_{-2100}$ in solar units \citep{Kirby2015}.  As for Draco II, these estimates (together with very low metallicity, $[\rm{Fe/H}]=-2.5\pm0.8$) suggest that Triangulum II is a dwarf galaxy. A later analysis by \citealp{Martin2} reported somehow different results ($\sigma_{vr} = 9.9^{+3.2}_{-2.2}$ ${\rm km\, s^{-1}}$, $M_{1/2}\sim 3\times10^6$\Msun, $(M/L)_{1/2}\sim15500$) with inner and outer stars' velocities in slight disagreement and with the nature of Triangulum II more uncertain. With current data, the uncertainty on these numbers is too big to have statistical significance and therefore we choose to continue our analysis assuming Triangulum II is a faint dwarf galaxy and using the estimates in \citealp{Kirby2015}. Both analyses, however, highlight that Triangulum II has a very large mass-to-light ratio, classifying it as the most DM dominated dwarf galaxy ever observed. 

\end{itemize}
\section{m22 Estimate}
\label{sec:m22}
To estimate the DM particle mass we match enclosed mass predictions with the observed half-light masses; fitting theoretical haloes to half-light measurements of satellite galaxies has been widely done in the literature to investigate different density profiles (see e.g., \citealp{Collins2014,BDC2015}). 

\citealp{Schive1} and \citealp{MPop2015} constrained the DM particle mass performing a simplified Jeans analysis on the resolved stellar populations of Fornax and analyzing the mass profile slopes of Fornax and Sculptor dwarfs, respectively. A full Jeans analysis using the standard 8 Milky Way dwarf satellite galaxies to estimate $m_{22}$ (which is indeed a universal parameter) will soon appear in \citealp{GM2016}. The galaxies we consider here are very faint and observations do not have the needed resolution to better resolve the stellar dynamics (e.g., no information on velocity anisotropy). Our calculations are then preliminary estimates of the DM component and we defer a more extensive analysis to future data.\\

We use two observational limits:
\begin{itemize}
\item The half-light mass, $M_{1/2}$\\
We assume that the stellar populations observed in Draco II and Triangulum II are within the respective core radius and we use the $M_{1/2}$ measurements to anchor the enclosed mass profile:
\be
M_{1/2}=\int_0^{r_{1/2}} \frac{4\pi 1.9(10 m_{22})^{-2}r_c^{-4}r'^2 }{[1+9.1\times10^{-2}(r'/r_c)^2]^8} dr' 10^9  {\rm M}_{\odot} \, ,
\label{eq:m12}
\ee
where $r_{1/2}=4/3r_h$ is the 3D deprojected half-light radius \citep{Wolf2010}.

This assumption is a good approximation with the current resolution, in both cases only very few stars are resolved (9 in Draco II and 6 in Triangulum II) and they seem to be in a single stellar system. \\

\item The maximum halo mass, $M_h^{\rm max}$\\
We impose a maximum halo mass based on the mass function of the Milky Way dwarf satellite galaxies \citep{Goccioli2008}. We choose  $M_h^{\rm max}$ to be $10^{-2}M_{\rm MW}\sim2\times10^{10} {\rm M}_{\odot}$. Such a maximum halo mass for satellites is in agreement with recent Local Group abundance matching results \citep{Brooks2014}. Moreover, $M_h^{\rm max}> {\rm few} \times 10^{10}$\Msun\, are forbidden by dynamical friction time scales limits \citep{Gerhard1992}. 

We fold this into Eq.~\ref{eq:rc} and get:
\be
m_{22}=\frac{1.6}{r_c}\Big(\frac{M_h^{\rm max}}{10^9{\rm M}_\odot}\Big)^{-1/3} \,.
\label{eq:m22}
\ee
We will discuss later how the results change had we chosen a different $M_h^{\rm max}$.
\end{itemize}

\subsection{Numerical results}
We can now combine Eq.~\ref{eq:m12} and Eq.~\ref{eq:m22} and numerically estimate the two parameters of the model ($r_c$ - $m_{22}$) for the two galaxies. We find:
\be r_c\sim105\, {\rm pc}; \, m_{22}\sim5.6 \,
\ee
for Draco II and 
\be 
r_c\sim160\, {\rm pc}; \, m_{22}\sim3.8
\ee
for Triangulum II. 

The corresponding core masses, using Eq.~\ref{eq:mc}, are $M_c^{\rm DraII}\sim 1.5\times 10^7$\Msun\, and $M_c^{\rm TriII}\sim 2.3\times 10^7$\Msun. 

\begin{figure}
\centering
\includegraphics[width=\columnwidth]{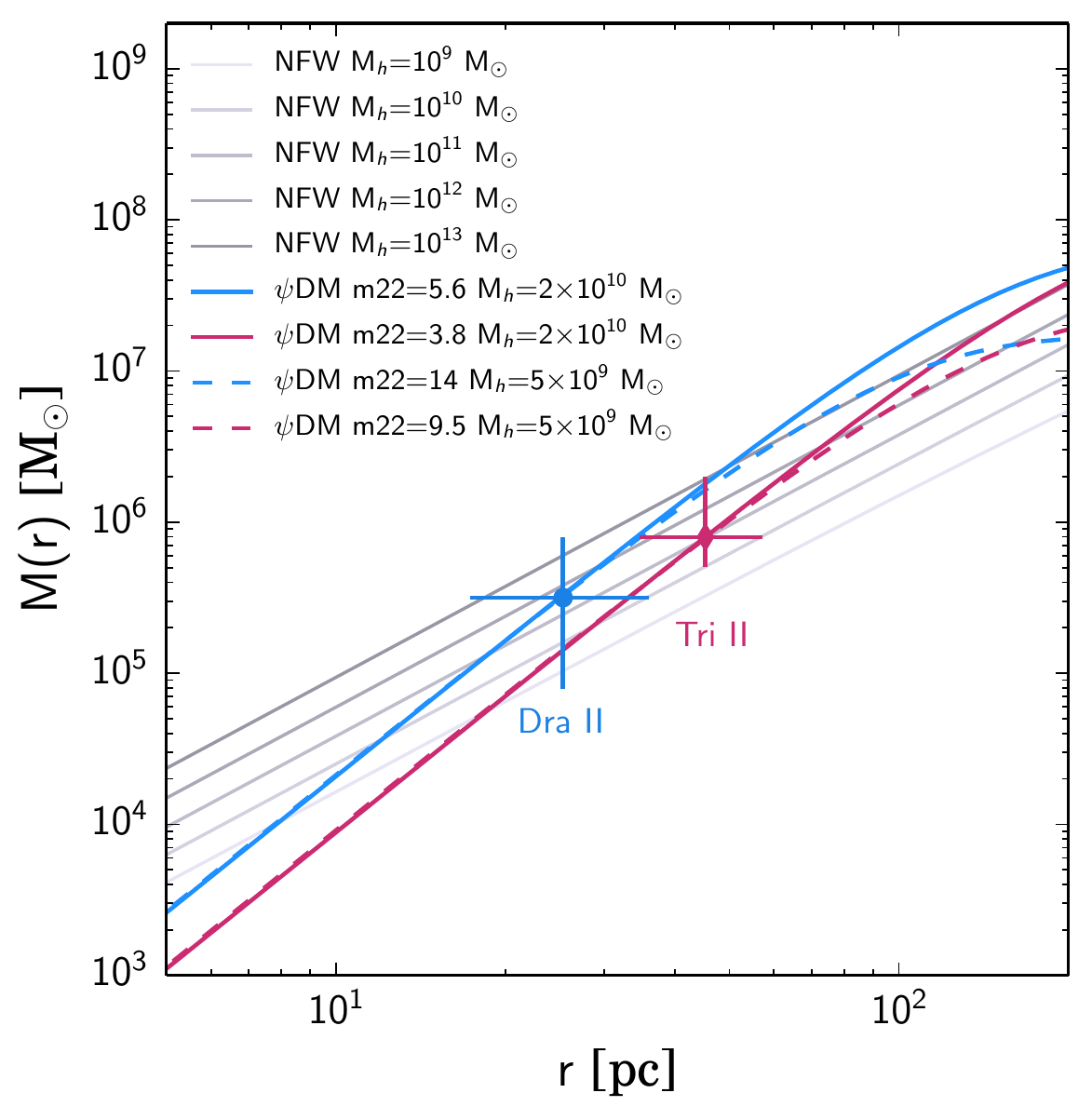}
\caption{Half-light mass measurements at the 3D deprojected half-light radius for Draco II \citep{Martin2015} and Triangulum II \citep{Kirby2015}. Grey lines, from lighter to darker, show predictions for the enclosed mass in a $\Lambda$CDM NFW model with M$_h$ going from $10^9$ to $10^{13}$\Msun. The NFW profiles are built using the scaling relations between the concentration parameter and the halo mass derived in \citealp{Maccio2008} and updated in \citealp{Dutton2014} for the \textit{Planck} 2013 cosmology \citep{PlanckParams2013} (Eq. 8 of \citealp{Dutton2014}). Solid colored lines show the estimated enclosed mass for the $\psi$DM model with $r_c\sim105 {\rm pc}$ - $m_{22}\sim5.6$ for Draco II and $r_c\sim160{\rm pc}$ - $m_{22}\sim3.8$ for Triangulum II. For the $\psi$DM model we compute only the soliton mass as total enclosed mass, this is a good approximation assuming that the stellar populations are within the core and that the soliton extends to $\sim3$ core radii \citep{Schive1,Schive2}. As a comparison, in dashed colored lines, we report $\psi$DM estimates for a less massive hosting halo; in the $\psi$DM model we can easily reconcile the data with smaller haloes at the expenses of a more massive DM particle.
\label{fig:m12r}}
\end{figure}

We show the $\psi$DM model predictions using these estimates and the half-light mass data in Fig.~\ref{fig:m12r}. The plot also shows NFW predictions for different values of the halo mass. In a NFW model these ultra-faint dwarfs lies systematically high compared to what we would expect, $M_h \approx 10^{8}-10^{10}$\Msun\, (see also Fig. 6 in \citealp{Martin2}). However, given the uncertainty on these measurements, the data points seem to agree with a wide range of values for the halo mass; all models with $M_h \approx 10^{9}-10^{13}$\Msun\, are within the $1-\sigma$ errorbar and no definitive conclusion can be drawn. This plot is in fact highlighting an intrinsic limitation of constraining mass profiles with the innermost regions of galaxies. As shown in \citealp{Ferrero2012}, different theoretical predictions lie very close to each other at very small radii ($r<1$kpc) and a wide range of models agree with the data. To break the models degeneracy, observations of stars outside the core, constraining the velocity dispersion -- or equivalently the mass -- at larger radii are needed. Nevertheless, the higher NFW curves conflict with physical constraints, $M_h$ can't reach the same mass of the host galaxy (the Milky Way in this case) and can't be bigger than the limits imposed by dynamical friction. In the case of $\psi$DM, we avoid this by imposing in the parameters model extraction the $M_h^{\rm max}$ quantity and finding a lower bound on $m_{22}$ corresponding to the maximum mass we allow for the halo. If we choose a lower value for $M_h^{\rm max}$, e.g., $\sim 5\times 10^9$\Msun, the estimates become $r_c\sim 66{\rm pc}$ - $m_{22}\sim14$ - $M_c\sim 3.8\times 10^6$\Msun\ for Draco II and $r_c\sim 99{\rm pc}$ - $m_{22}\sim9.5$ - $M_c\sim 5.7\times 10^6$\Msun\ for Triangulum II. This means that our estimates for $m_{22}$ in the case $M_h^{\rm max}=2\times 10^{10}$\Msun\, are lower limits for the DM particle mass. In the $\psi$DM model we can easily reconcile the data with smaller haloes at the expenses of a more massive DM particle. 

We want to stress that in these results we are strongly dominated by the anti-correlation between $r_c$ and $m_{22}$ (see Fig. S4 in \citealp{Schive1}), deeper observations, resolving multiple stellar systems positioned at different radii in the enclosed mass predictions, will be fundamental to get more robust estimates. 

\begin{figure}
\centering
\includegraphics[width=\columnwidth]{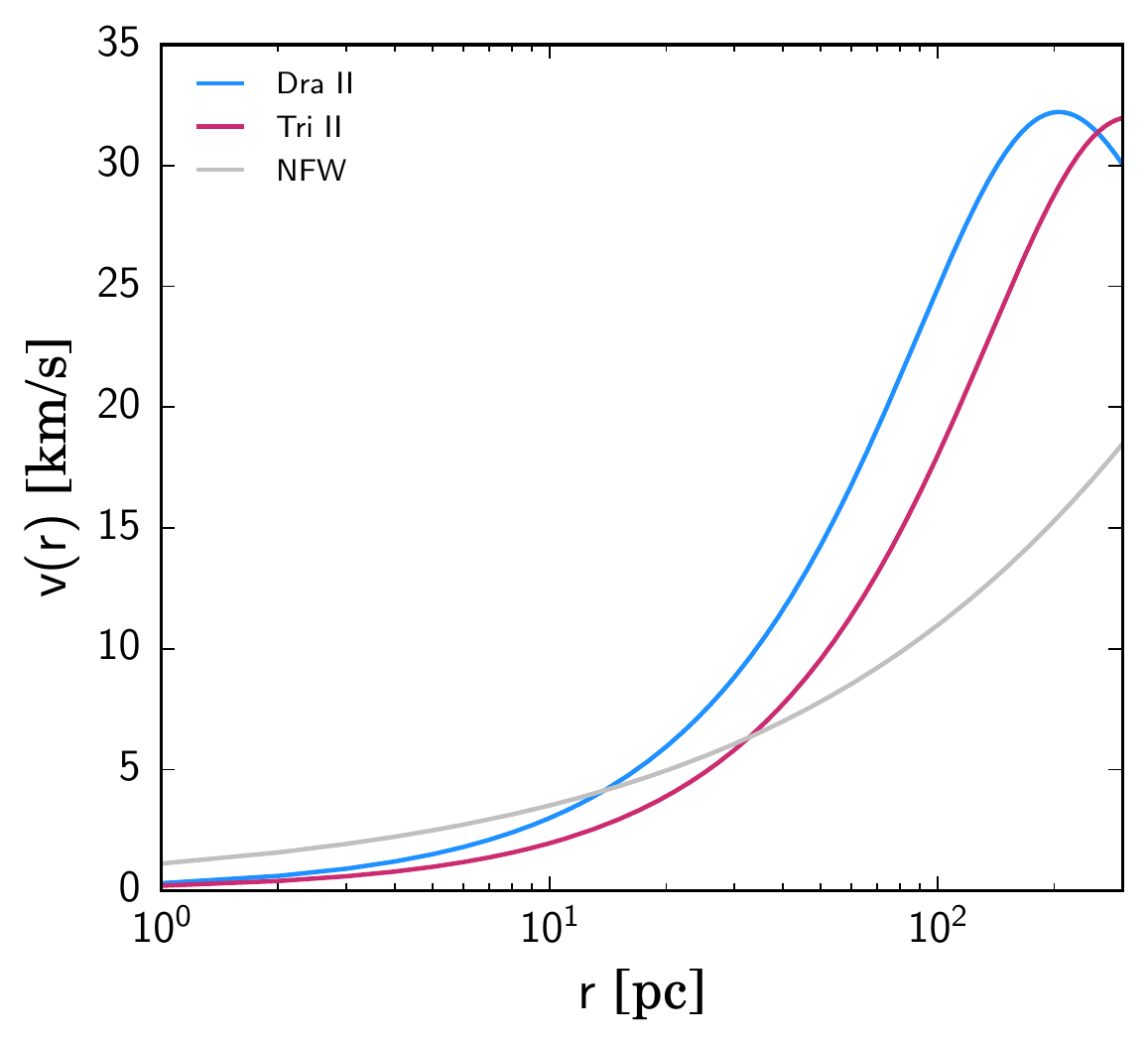}
\caption{Circular velocity distribution predicted using the $\psi$DM model estimates for Draco II (blue line) and Triangulum II (magenta line) and using a NFW density profile with a $2\times 10^{10}$\Msun\ halo mass.
\label{fig:vr}}
\end{figure}

We report in Fig.~\ref{fig:vr} the predicted circular velocities, $v(r)=\sqrt(GM(<r)/r)$, out to 3-times the core radius -- where the soliton density is a good approximation of the total halo density -- for the $\psi$DM estimates derived above and for a NFW profile with the same $M_h^{\rm max}$. As previously anticipated, the plot highlights that measurements of the stellar velocities at larger distances from the center will help distinguish between different curves and impose strong constraints on the density profiles. We note that this plot does not extend to the constant velocity regime generated by the additional NFW-like density profile term overtaking the soliton profile at large scales.

\section{discussion}
In this paper we have estimated the dark matter particle mass from newly discovered ultra-faint dwarf galaxies in a scenario where dark matter is made of ultra-light bosons, axion-like particles, condensate in a coherent wave. 

Wave Dark Matter ($\psi$DM) has recently been a very valuable alternative to Cold Dark Matter (CDM) to solve the small-scale crisis of CDM in galaxy formation \citep{Schive1,Schive2,Schive3,MPop2015}. The $\psi$DM model infact superpose a soliton core depending on the DM particle mass $m$ on a  Navarro-Frenk-White (NFW) density profile, smoothing the unwanted cusp in the center of the density profile and suppressing formation of haloes smaller than $10^{10}$\Msun. At large scales $\psi$DM clustering is statistically indistinguishable from CDM.

We have considered in this work data from recently discovered Draco II and Triangulum II Milky Way dwarf satellites. The measurements of the luminosity of these objects suggest that they are systems with a mass-to-light ratio of $\sim 10^7$ in the outer regions and therefore completely DM dominated. With such extreme DM environments, these dwarfs provide a great laboratory to study dark matter physics despite the contribution of baryonic feedback.

We anchor the soliton density profile of the $\psi$DM model using Draco II and Triangulum II half-light mass measurements and limits from the halo mass function of Milky Way satellite galaxies, and estimate at the same time the core radius $r_c$ and the DM particle mass $m$. We find that the $\psi$DM model requires $r_c\sim105{\rm pc}$ - $m\sim5.6\times 10^{-22}{\rm eV}$ for Draco II and $r_c\sim160{\rm pc}$ - $m\sim3.8\times 10^{-22}{\rm eV}$ for Triangulum II. With these values of the parameters, the haloes hosting the galaxies have a total mass of $2\times 10^{10}$\Msun. If we impose a smaller halo mass, we can readjust the estimates of $r_c$ and $m$ and embed these dwarfs in less massive haloes with a smaller core radius and a more massive DM particle. 

Ultra-light axion-like particles have been recently tested with many cosmological probes as an alternative to CDM and the interesting scenario of many data supporting $10^{-23}{\rm eV} \lesssim m \lesssim 10^{-21}{\rm eV}$ is lately emerging: Lyman-$\alpha$ systems constraints are presented in \citealp{Amendola2006}, galaxy formation, galaxy dynamics and UV luminosity function tests in \citealp{Lora2014,Lora2015,Schive1,Schive3,MPop2015,Marsh2013}, reionization constraints in \citealp{Bozek2014,Sarkar2015} and Cosmic Microwave Background limits in \citealp{Hlozek2014}. Our estimates fit very nicely into these bounds and highlight how promising ultra-faint dwarf galaxies are in testing DM models. 

Deeper observations from future experiments like JWST \citep{JWST}, WFRIST \citep{WFIRST}, LSST \citep{LSST}, and Euclid \citep{Euclid} will enable further discoveries of Milky Way satellites and will characterize the stellar systems within these dwarfs with the resolution needed for more detailed analysis. 
They will, for example, enable some potential tests for the $\psi$DM model:
\begin{itemize}[noitemsep,nolistsep]
\item{\textit{minimum halo mass}} --
Following \citealp{Schive1,Schive2} we can estimate the minimum halo mass possible in this scenario, $M_h^{\rm min}\sim4.4\times10^7 m_{22}^{-3/2}\sim 5\times 10^6$\Msun\,. These objects will be the target of sub-haloes detection via milli-lensing \citep{Dalal2002} and strong-lensing \citep{Hezaveh2014}.
\item{\textit{late galaxy formation}} -- Lyman-$\alpha$ emission, galaxy luminosity function (from JWST) and reionization (from CMB \citep{Calabrese2014} or 21cm \citep{Kadota2014}) will test the suppression of galaxy formation at high redshift predicted by this model \citep{Marsh2013,Schive3}. 
\item{\textit{sub-halo masses}} -- The halo mass function will manifest a truncation at low masses in the case of $\psi$DM \citep{Marsh2013, Schive3}, with a suppression of objects below the Jeans scale set by the DM particle mass. Moreover, considering that our knowledge of the Galactic plane will soon dramatically improve with Gaia \citep{Gaia}, we can test 
this natural cut-off of the model by observing the predicted suppression of tidal streams. 
\end{itemize}

\section*{Acknowledgments}
We are grateful to D.J.E. Marsh and H.Y. Schive for useful discussions. We especially thank D.J.E. Marsh, J.P. Ostriker, A. Di Cintio and R. Hlozek for reading the manuscript and providing useful feedback. EC is supported by Lyman Spitzer Jr. and STFC Rutherford Fellowships.
\label{sec:disc}
\bibliographystyle{mnras}
\bibliography{ref}

\begin{thebibliography}{}
\makeatletter
\relax
\def\mn@urlcharsother{\let\do\@makeother \do\$\do\&\do\#\do\^\do\_\do\%\do\~}
\def\mn@doi{\begingroup\mn@urlcharsother \@ifnextchar [ {\mn@doi@}
  {\mn@doi@[]}}
\def\mn@doi@[#1]#2{\def\@tempa{#1}\ifx\@tempa\@empty \href
  {http://dx.doi.org/#2} {doi:#2}\else \href {http://dx.doi.org/#2} {#1}\fi
  \endgroup}
\def\mn@eprint#1#2{\mn@eprint@#1:#2::\@nil}
\def\mn@eprint@arXiv#1{\href {http://arxiv.org/abs/#1} {{\tt arXiv:#1}}}
\def\mn@eprint@dblp#1{\href {http://dblp.uni-trier.de/rec/bibtex/#1.xml}
  {dblp:#1}}
\def\mn@eprint@#1:#2:#3:#4\@nil{\def\@tempa {#1}\def\@tempb {#2}\def\@tempc
  {#3}\ifx \@tempc \@empty \let \@tempc \@tempb \let \@tempb \@tempa \fi \ifx
  \@tempb \@empty \def\@tempb {arXiv}\fi \@ifundefined
  {mn@eprint@\@tempb}{\@tempb:\@tempc}{\expandafter \expandafter \csname
  mn@eprint@\@tempb\endcsname \expandafter{\@tempc}}}

\bibitem[\protect\citeauthoryear{{Amendola} \& {Barbieri}}{{Amendola} \&
  {Barbieri}}{2006}]{Amendola2006}
{Amendola} L.,  {Barbieri} R.,  2006, \mn@doi [Physics Letters B]
  {10.1016/j.physletb.2006.08.069}, \href
  {http://adsabs.harvard.edu/abs/2006PhLB..642..192A} {642, 192}

\bibitem[\protect\citeauthoryear{{Bechtol} et~al.,}{{Bechtol}
  et~al.}{2015}]{Des1}
{Bechtol} K.,  et~al., 2015, \mn@doi [\apj] {10.1088/0004-637X/807/1/50}, \href
  {http://adsabs.harvard.edu/abs/2015ApJ...807...50B} {807, 50}

\bibitem[\protect\citeauthoryear{{Belokurov} et~al.,}{{Belokurov}
  et~al.}{2007}]{Belokurov2007}
{Belokurov} V.,  et~al., 2007, \mn@doi [\apj] {10.1086/509718}, \href
  {http://adsabs.harvard.edu/abs/2007ApJ...654..897B} {654, 897}

\bibitem[\protect\citeauthoryear{{Blumenthal}, {Faber}, {Primack}  \&
  {Rees}}{{Blumenthal} et~al.}{1984}]{Blum84}
{Blumenthal} G.~R.,  {Faber} S.~M.,  {Primack} J.~R.,   {Rees} M.~J.,  1984,
  \mn@doi [\nat] {10.1038/311517a0}, \href
  {http://adsabs.harvard.edu/abs/1984Natur.311..517B} {311, 517}

\bibitem[\protect\citeauthoryear{{B{\"o}hmer} \& {Harko}}{{B{\"o}hmer} \&
  {Harko}}{2007}]{Bohmer2007}
{B{\"o}hmer} C.~G.,  {Harko} T.,  2007, \mn@doi [\jcap]
  {10.1088/1475-7516/2007/06/025}, \href
  {http://adsabs.harvard.edu/abs/2007JCAP...06..025B} {6, 25}

\bibitem[\protect\citeauthoryear{{Bozek}, {Marsh}, {Silk}  \& {Wyse}}{{Bozek}
  et~al.}{2015}]{Bozek2014}
{Bozek} B.,  {Marsh} D.~J.~E.,  {Silk} J.,   {Wyse} R.~F.~G.,  2015, \mn@doi
  [\mnras] {10.1093/mnras/stv624}, \href
  {http://adsabs.harvard.edu/abs/2015MNRAS.450..209B} {450, 209}

\bibitem[\protect\citeauthoryear{{Brook} \& {Di Cintio}}{{Brook} \& {Di
  Cintio}}{2015}]{BDC2015}
{Brook} C.~B.,  {Di Cintio} A.,  2015, \mn@doi [\mnras] {10.1093/mnras/stv864},
  \href {http://adsabs.harvard.edu/abs/2015MNRAS.450.3920B} {450, 3920}

\bibitem[\protect\citeauthoryear{{Brook}, {Di Cintio}, {Knebe},
  {Gottl{\"o}ber}, {Hoffman}, {Yepes}  \& {Garrison-Kimmel}}{{Brook}
  et~al.}{2014}]{Brooks2014}
{Brook} C.~B.,  {Di Cintio} A.,  {Knebe} A.,  {Gottl{\"o}ber} S.,  {Hoffman}
  Y.,  {Yepes} G.,   {Garrison-Kimmel} S.,  2014, \mn@doi [\apjl]
  {10.1088/2041-8205/784/1/L14}, \href
  {http://adsabs.harvard.edu/abs/2014ApJ...784L..14B} {784, L14}

\bibitem[\protect\citeauthoryear{{Calabrese} et~al.,}{{Calabrese}
  et~al.}{2014}]{Calabrese2014}
{Calabrese} E.,  et~al., 2014, \mn@doi [\jcap] {10.1088/1475-7516/2014/08/010},
  \href {http://adsabs.harvard.edu/abs/2014JCAP...08..010C} {8, 010}

\bibitem[\protect\citeauthoryear{{Chavanis}}{{Chavanis}}{2011}]{Chavanis2011}
{Chavanis} P.-H.,  2011, \mn@doi [\prd] {10.1103/PhysRevD.84.043531}, \href
  {http://adsabs.harvard.edu/abs/2011PhRvD..84d3531C} {84, 043531}

\bibitem[\protect\citeauthoryear{{Chavanis}}{{Chavanis}}{2012}]{Chavanis2012}
{Chavanis} P.~H.,  2012, \mn@doi [\aap] {10.1051/0004-6361/201116905}, \href
  {http://adsabs.harvard.edu/abs/2012A%26A...537A.127C} {537, A127}

\bibitem[\protect\citeauthoryear{{Chavanis} \& {Delfini}}{{Chavanis} \&
  {Delfini}}{2011}]{Chavanis2}
{Chavanis} P.-H.,  {Delfini} L.,  2011, \mn@doi [\prd]
  {10.1103/PhysRevD.84.043532}, \href
  {http://adsabs.harvard.edu/abs/2011PhRvD..84d3532C} {84, 043532}

\bibitem[\protect\citeauthoryear{{Collins} et~al.,}{{Collins}
  et~al.}{2014}]{Collins2014}
{Collins} M.~L.~M.,  et~al., 2014, \mn@doi [\apj] {10.1088/0004-637X/783/1/7},
  \href {http://adsabs.harvard.edu/abs/2014ApJ...783....7C} {783, 7}

\bibitem[\protect\citeauthoryear{{Dalal} \& {Kochanek}}{{Dalal} \&
  {Kochanek}}{2002}]{Dalal2002}
{Dalal} N.,  {Kochanek} C.~S.,  2002, \mn@doi [\apj] {10.1086/340303}, \href
  {http://adsabs.harvard.edu/abs/2002ApJ...572...25D} {572, 25}

\bibitem[\protect\citeauthoryear{{Davidson}}{{Davidson}}{2015}]{Davidson2015}
{Davidson} S.,  2015, \mn@doi [Astroparticle Physics]
  {10.1016/j.astropartphys.2014.12.007}, \href
  {http://adsabs.harvard.edu/abs/2015APh....65..101D} {65, 101}

\bibitem[\protect\citeauthoryear{{Di Cintio}, {Brook}, {Macci{\`o}}, {Stinson},
  {Knebe}, {Dutton}  \& {Wadsley}}{{Di Cintio} et~al.}{2014}]{DiCintio2014}
{Di Cintio} A.,  {Brook} C.~B.,  {Macci{\`o}} A.~V.,  {Stinson} G.~S.,  {Knebe}
  A.,  {Dutton} A.~A.,   {Wadsley} J.,  2014, \mn@doi [\mnras]
  {10.1093/mnras/stt1891}, \href
  {http://adsabs.harvard.edu/abs/2014MNRAS.437..415D} {437, 415}

\bibitem[\protect\citeauthoryear{{Drlica-Wagner} et~al.,}{{Drlica-Wagner}
  et~al.}{2015}]{Des2}
{Drlica-Wagner} A.,  et~al., 2015, \mn@doi [\apj]
  {10.1088/0004-637X/813/2/109}, \href
  {http://adsabs.harvard.edu/abs/2015ApJ...813..109D} {813, 109}

\bibitem[\protect\citeauthoryear{{Dutton} \& {Macci{\`o}}}{{Dutton} \&
  {Macci{\`o}}}{2014}]{Dutton2014}
{Dutton} A.~A.,  {Macci{\`o}} A.~V.,  2014, \mn@doi [\mnras]
  {10.1093/mnras/stu742}, \href
  {http://adsabs.harvard.edu/abs/2014MNRAS.441.3359D} {441, 3359}

\bibitem[\protect\citeauthoryear{{Ferrero}, {Abadi}, {Navarro}, {Sales}  \&
  {Gurovich}}{{Ferrero} et~al.}{2012}]{Ferrero2012}
{Ferrero} I.,  {Abadi} M.~G.,  {Navarro} J.~F.,  {Sales} L.~V.,   {Gurovich}
  S.,  2012, \mn@doi [\mnras] {10.1111/j.1365-2966.2012.21623.x}, \href
  {http://adsabs.harvard.edu/abs/2012MNRAS.425.2817F} {425, 2817}

\bibitem[\protect\citeauthoryear{{Gardner} et~al.,}{{Gardner}
  et~al.}{2006}]{JWST}
{Gardner} J.~P.,  et~al., 2006, \mn@doi [\ssr] {10.1007/s11214-006-8315-7},
  \href {http://adsabs.harvard.edu/abs/2006SSRv..123..485G} {123, 485}

\bibitem[\protect\citeauthoryear{{Gerhard} \& {Spergel}}{{Gerhard} \&
  {Spergel}}{1992}]{Gerhard1992}
{Gerhard} O.~E.,  {Spergel} D.~N.,  1992, \mn@doi [\apjl] {10.1086/186336},
  \href {http://adsabs.harvard.edu/abs/1992ApJ...389L...9G} {389, L9}

\bibitem[\protect\citeauthoryear{{Giocoli}, {Tormen}  \& {van den
  Bosch}}{{Giocoli} et~al.}{2008}]{Goccioli2008}
{Giocoli} C.,  {Tormen} G.,   {van den Bosch} F.~C.,  2008, \mn@doi [\mnras]
  {10.1111/j.1365-2966.2008.13182.x}, \href
  {http://adsabs.harvard.edu/abs/2008MNRAS.386.2135G} {386, 2135}

\bibitem[\protect\citeauthoryear{{Gonzalez-Morales}, {Marsh}, {Penarrubia}  \&
  {Urena-Lopez}}{{Gonzalez-Morales} et~al.}{2016}]{GM2016}
{Gonzalez-Morales} A.~X.,  {Marsh} D.~J.~E.,  {Penarrubia} J.,   {Urena-Lopez}
  L.,  2016, In prep.

\bibitem[\protect\citeauthoryear{{Goodman}}{{Goodman}}{2000}]{Goodman2000}
{Goodman} J.,  2000, \mn@doi [\na] {10.1016/S1384-1076(00)00015-4}, \href
  {http://adsabs.harvard.edu/abs/2000NewA....5..103G} {5, 103}

\bibitem[\protect\citeauthoryear{{Governato} et~al.,}{{Governato}
  et~al.}{2012}]{Governato2012}
{Governato} F.,  et~al., 2012, \mn@doi [\mnras]
  {10.1111/j.1365-2966.2012.20696.x}, \href
  {http://adsabs.harvard.edu/abs/2012MNRAS.422.1231G} {422, 1231}

\bibitem[\protect\citeauthoryear{{Guth}, {Hertzberg}  \&
  {Prescod-Weinstein}}{{Guth} et~al.}{2015}]{Guth2015}
{Guth} A.~H.,  {Hertzberg} M.~P.,   {Prescod-Weinstein} C.,  2015, \mn@doi
  [\prd] {10.1103/PhysRevD.92.103513}, \href
  {http://adsabs.harvard.edu/abs/2015PhRvD..92j3513G} {92, 103513}

\bibitem[\protect\citeauthoryear{{Guzm{\'a}n} \& {Lora-Clavijo}}{{Guzm{\'a}n}
  \& {Lora-Clavijo}}{2015}]{Guzman2015}
{Guzm{\'a}n} F.~S.,  {Lora-Clavijo} F.~D.,  2015, \mn@doi [General Relativity
  and Gravitation] {10.1007/s10714-015-1865-9}, \href
  {http://adsabs.harvard.edu/abs/2015GReGr..47...21G} {47, 21}

\bibitem[\protect\citeauthoryear{{Guzm{\'a}n}, {Matos}  \&
  {Villegas}}{{Guzm{\'a}n} et~al.}{1999}]{Guzman1999}
{Guzm{\'a}n} F.~S.,  {Matos} T.,   {Villegas} H.~B.,  1999, Astronomische
  Nachrichten, \href {http://adsabs.harvard.edu/abs/1999AN....320...97G} {320,
  97}

\bibitem[\protect\citeauthoryear{{Harko}}{{Harko}}{2011a}]{Harko20112}
{Harko} T.,  2011a, \mn@doi [\jcap] {10.1088/1475-7516/2011/05/022}, \href
  {http://adsabs.harvard.edu/abs/2011JCAP...05..022H} {5, 22}

\bibitem[\protect\citeauthoryear{{Harko}}{{Harko}}{2011b}]{Harko2011}
{Harko} T.,  2011b, \mn@doi [\prd] {10.1103/PhysRevD.83.123515}, \href
  {http://adsabs.harvard.edu/abs/2011PhRvD..83l3515H} {83, 123515}

\bibitem[\protect\citeauthoryear{{Harko}}{{Harko}}{2014}]{Harko2014}
{Harko} T.,  2014, \mn@doi [\prd] {10.1103/PhysRevD.89.084040}, \href
  {http://adsabs.harvard.edu/abs/2014PhRvD..89h4040H} {89, 084040}

\bibitem[\protect\citeauthoryear{{Harko} \& {Lobo}}{{Harko} \&
  {Lobo}}{2015}]{Harko20152}
{Harko} T.,  {Lobo} F.~S.~N.,  2015, \mn@doi [\prd]
  {10.1103/PhysRevD.92.043011}, \href
  {http://adsabs.harvard.edu/abs/2015PhRvD..92d3011H} {92, 043011}

\bibitem[\protect\citeauthoryear{{Harko} \& {Madarassy}}{{Harko} \&
  {Madarassy}}{2012}]{Harko20122}
{Harko} T.,  {Madarassy} E.~J.~M.,  2012, \mn@doi [\jcap]
  {10.1088/1475-7516/2012/01/020}, \href
  {http://adsabs.harvard.edu/abs/2012JCAP...01..020H} {1, 20}

\bibitem[\protect\citeauthoryear{{Harko} \& {Mocanu}}{{Harko} \&
  {Mocanu}}{2012}]{Harko2012}
{Harko} T.,  {Mocanu} G.,  2012, \mn@doi [\prd] {10.1103/PhysRevD.85.084012},
  \href {http://adsabs.harvard.edu/abs/2012PhRvD..85h4012H} {85, 084012}

\bibitem[\protect\citeauthoryear{{Harko}, {Liang}, {Liang}  \&
  {Mocanu}}{{Harko} et~al.}{2015}]{Harko2015}
{Harko} T.,  {Liang} P.,  {Liang} S.-D.,   {Mocanu} G.,  2015, \mn@doi [\jcap]
  {10.1088/1475-7516/2015/11/027}, \href
  {http://adsabs.harvard.edu/abs/2015JCAP...11..027H} {11, 27}

\bibitem[\protect\citeauthoryear{{Hezaveh}, {Dalal}, {Holder}, {Kisner},
  {Kuhlen}  \& {Perreault Levasseur}}{{Hezaveh} et~al.}{2014}]{Hezaveh2014}
{Hezaveh} Y.,  {Dalal} N.,  {Holder} G.,  {Kisner} T.,  {Kuhlen} M.,
  {Perreault Levasseur} L.,  2014, preprint, \href
  {http://adsabs.harvard.edu/abs/2014arXiv1403.2720H} {} (\mn@eprint {arXiv}
  {1403.2720})

\bibitem[\protect\citeauthoryear{{Hlozek}, {Grin}, {Marsh}  \&
  {Ferreira}}{{Hlozek} et~al.}{2015}]{Hlozek2014}
{Hlozek} R.,  {Grin} D.,  {Marsh} D.~J.~E.,   {Ferreira} P.~G.,  2015, \mn@doi
  [\prd] {10.1103/PhysRevD.91.103512}, \href
  {http://adsabs.harvard.edu/abs/2015PhRvD..91j3512H} {91, 103512}

\bibitem[\protect\citeauthoryear{{Hu}, {Barkana}  \& {Gruzinov}}{{Hu}
  et~al.}{2000}]{Hu00}
{Hu} W.,  {Barkana} R.,   {Gruzinov} A.,  2000, \mn@doi [Physical Review
  Letters] {10.1103/PhysRevLett.85.1158}, \href
  {http://adsabs.harvard.edu/abs/2000PhRvL..85.1158H} {85, 1158}

\bibitem[\protect\citeauthoryear{{Ji} \& {Sin}}{{Ji} \& {Sin}}{1994}]{Ji1994}
{Ji} S.~U.,  {Sin} S.~J.,  1994, \mn@doi [\prd] {10.1103/PhysRevD.50.3655},
  \href {http://adsabs.harvard.edu/abs/1994PhRvD..50.3655J} {50, 3655}

\bibitem[\protect\citeauthoryear{{Kadota}, {Mao}, {Ichiki}  \& {Silk}}{{Kadota}
  et~al.}{2014}]{Kadota2014}
{Kadota} K.,  {Mao} Y.,  {Ichiki} K.,   {Silk} J.,  2014, \mn@doi [\jcap]
  {10.1088/1475-7516/2014/06/011}, \href
  {http://adsabs.harvard.edu/abs/2014JCAP...06..011K} {6, 011}

\bibitem[\protect\citeauthoryear{{Kirby}, {Cohen}, {Simon}  \&
  {Guhathakurta}}{{Kirby} et~al.}{2015}]{Kirby2015}
{Kirby} E.~N.,  {Cohen} J.~G.,  {Simon} J.~D.,   {Guhathakurta} P.,  2015,
  \mn@doi [\apjl] {10.1088/2041-8205/814/1/L7}, \href
  {http://adsabs.harvard.edu/abs/2015ApJ...814L...7K} {814, L7}

\bibitem[\protect\citeauthoryear{{Koposov}, {Belokurov}, {Torrealba}  \&
  {Evans}}{{Koposov} et~al.}{2015}]{Koposov2015}
{Koposov} S.~E.,  {Belokurov} V.,  {Torrealba} G.,   {Evans} N.~W.,  2015,
  \mn@doi [\apj] {10.1088/0004-637X/805/2/130}, \href
  {http://adsabs.harvard.edu/abs/2015ApJ...805..130K} {805, 130}

\bibitem[\protect\citeauthoryear{{Kuhlen}, {Vogelsberger}  \&
  {Angulo}}{{Kuhlen} et~al.}{2012}]{Kuhlen2012}
{Kuhlen} M.,  {Vogelsberger} M.,   {Angulo} R.,  2012, \mn@doi [Physics of the
  Dark Universe] {10.1016/j.dark.2012.10.002}, \href
  {http://adsabs.harvard.edu/abs/2012PDU.....1...50K} {1, 50}

\bibitem[\protect\citeauthoryear{{LSST Science Collaboration} et~al.,}{{LSST
  Science Collaboration} et~al.}{2009}]{LSST}
{LSST Science Collaboration} et~al., 2009, preprint, \href
  {http://adsabs.harvard.edu/abs/2009arXiv0912.0201L} {} (\mn@eprint {arXiv}
  {0912.0201})

\bibitem[\protect\citeauthoryear{{Laevens} et~al.,}{{Laevens}
  et~al.}{2015a}]{Laevens20150}
{Laevens} B.~P.~M.,  et~al., 2015a, \mn@doi [\apjl]
  {10.1088/2041-8205/802/2/L18}, \href
  {http://adsabs.harvard.edu/abs/2015ApJ...802L..18L} {802, L18}

\bibitem[\protect\citeauthoryear{{Laevens} et~al.,}{{Laevens}
  et~al.}{2015b}]{Laevens2015}
{Laevens} B.~P.~M.,  et~al., 2015b, \mn@doi [\apj]
  {10.1088/0004-637X/813/1/44}, \href
  {http://adsabs.harvard.edu/abs/2015ApJ...813...44L} {813, 44}

\bibitem[\protect\citeauthoryear{{Laureijs} et~al.,}{{Laureijs}
  et~al.}{2011}]{Euclid}
{Laureijs} R.,  et~al., 2011, preprint, \href
  {http://adsabs.harvard.edu/abs/2011arXiv1110.3193L} {} (\mn@eprint {arXiv}
  {1110.3193})

\bibitem[\protect\citeauthoryear{{Lee} \& {Koh}}{{Lee} \&
  {Koh}}{1996}]{Lee1996}
{Lee} J.-W.,  {Koh} I.-G.,  1996, \mn@doi [\prd] {10.1103/PhysRevD.53.2236},
  \href {http://adsabs.harvard.edu/abs/1996PhRvD..53.2236L} {53, 2236}

\bibitem[\protect\citeauthoryear{{Li}, {Rindler-Daller}  \& {Shapiro}}{{Li}
  et~al.}{2014}]{Li2014}
{Li} B.,  {Rindler-Daller} T.,   {Shapiro} P.~R.,  2014, \mn@doi [\prd]
  {10.1103/PhysRevD.89.083536}, \href
  {http://adsabs.harvard.edu/abs/2014PhRvD..89h3536L} {89, 083536}

\bibitem[\protect\citeauthoryear{{Lora}}{{Lora}}{2015}]{Lora2015}
{Lora} V.,  2015, \mn@doi [\apj] {10.1088/0004-637X/807/2/116}, \href
  {http://adsabs.harvard.edu/abs/2015ApJ...807..116L} {807, 116}

\bibitem[\protect\citeauthoryear{{Lora} \& {Maga{\~n}a}}{{Lora} \&
  {Maga{\~n}a}}{2014}]{Lora2014}
{Lora} V.,  {Maga{\~n}a} J.,  2014, \mn@doi [\jcap]
  {10.1088/1475-7516/2014/09/011}, \href
  {http://adsabs.harvard.edu/abs/2014JCAP...09..011L} {9, 11}

\bibitem[\protect\citeauthoryear{{Lora}, {Maga{\~n}a}, {Bernal},
  {S{\'a}nchez-Salcedo}  \& {Grebel}}{{Lora} et~al.}{2012}]{Lora2012}
{Lora} V.,  {Maga{\~n}a} J.,  {Bernal} A.,  {S{\'a}nchez-Salcedo} F.~J.,
  {Grebel} E.~K.,  2012, \mn@doi [\jcap] {10.1088/1475-7516/2012/02/011}, \href
  {http://adsabs.harvard.edu/abs/2012JCAP...02..011L} {2, 11}

\bibitem[\protect\citeauthoryear{{Lundgren}, {Bondarescu}, {Bondarescu}  \&
  {Balakrishna}}{{Lundgren} et~al.}{2010}]{Lundgren2010}
{Lundgren} A.~P.,  {Bondarescu} M.,  {Bondarescu} R.,   {Balakrishna} J.,
  2010, \mn@doi [\apjl] {10.1088/2041-8205/715/1/L35}, \href
  {http://adsabs.harvard.edu/abs/2010ApJ...715L..35L} {715, L35}

\bibitem[\protect\citeauthoryear{{Macci{\`o}}, {Dutton}  \& {van den
  Bosch}}{{Macci{\`o}} et~al.}{2008}]{Maccio2008}
{Macci{\`o}} A.~V.,  {Dutton} A.~A.,   {van den Bosch} F.~C.,  2008, \mn@doi
  [\mnras] {10.1111/j.1365-2966.2008.14029.x}, \href
  {http://adsabs.harvard.edu/abs/2008MNRAS.391.1940M} {391, 1940}

\bibitem[\protect\citeauthoryear{{Maga{\~n}a} \& {Matos}}{{Maga{\~n}a} \&
  {Matos}}{2012}]{Magana2012}
{Maga{\~n}a} J.,  {Matos} T.,  2012, \mn@doi [Journal of Physics Conference
  Series] {10.1088/1742-6596/378/1/012012}, \href
  {http://adsabs.harvard.edu/abs/2012JPhCS.378a2012M} {378, 012012}

\bibitem[\protect\citeauthoryear{{Marsh}}{{Marsh}}{2015a}]{Marsh}
{Marsh} D.~J.~E.,  2015a, preprint, \href
  {http://adsabs.harvard.edu/abs/2015arXiv151007633M} {} (\mn@eprint {arXiv}
  {1510.07633})

\bibitem[\protect\citeauthoryear{{Marsh}}{{Marsh}}{2015b}]{Marsh2}
{Marsh} D.~J.~E.,  2015b, \mn@doi [\prd] {10.1103/PhysRevD.91.123520}, \href
  {http://adsabs.harvard.edu/abs/2015PhRvD..91l3520M} {91, 123520}

\bibitem[\protect\citeauthoryear{{Marsh} \& {Pop}}{{Marsh} \&
  {Pop}}{2015}]{MPop2015}
{Marsh} D.~J.~E.,  {Pop} A.-R.,  2015, \mn@doi [\mnras]
  {10.1093/mnras/stv1050}, \href
  {http://adsabs.harvard.edu/abs/2015MNRAS.451.2479M} {451, 2479}

\bibitem[\protect\citeauthoryear{{Marsh} \& {Silk}}{{Marsh} \&
  {Silk}}{2014}]{Marsh2013}
{Marsh} D.~J.~E.,  {Silk} J.,  2014, \mn@doi [\mnras] {10.1093/mnras/stt2079},
  \href {http://adsabs.harvard.edu/abs/2014MNRAS.437.2652M} {437, 2652}

\bibitem[\protect\citeauthoryear{{Martin} et~al.,}{{Martin}
  et~al.}{2016a}]{Martin2015}
{Martin} N.~F.,  et~al., 2016a, \mn@doi [\mnras] {10.1093/mnrasl/slw013}, \href
  {http://adsabs.harvard.edu/abs/2016MNRAS.458L..59M} {458, L59}

\bibitem[\protect\citeauthoryear{{Martin} et~al.,}{{Martin}
  et~al.}{2016b}]{Martin2}
{Martin} N.~F.,  et~al., 2016b, \mn@doi [\apj] {10.3847/0004-637X/818/1/40},
  \href {http://adsabs.harvard.edu/abs/2016ApJ...818...40M} {818, 40}

\bibitem[\protect\citeauthoryear{{Matos}, {Guzm{\'a}n}  \&
  {Ure{\~n}a-L{\'o}pez}}{{Matos} et~al.}{2000}]{Guzman2000}
{Matos} T.,  {Guzm{\'a}n} F.~S.,   {Ure{\~n}a-L{\'o}pez} L.~A.,  2000, \mn@doi
  [Classical and Quantum Gravity] {10.1088/0264-9381/17/7/309}, \href
  {http://adsabs.harvard.edu/abs/2000CQGra..17.1707M} {17, 1707}

\bibitem[\protect\citeauthoryear{{Navarro}, {Frenk}  \& {White}}{{Navarro}
  et~al.}{1996}]{NFW}
{Navarro} J.~F.,  {Frenk} C.~S.,   {White} S.~D.~M.,  1996, \mn@doi [\apj]
  {10.1086/177173}, \href {http://adsabs.harvard.edu/abs/1996ApJ...462..563N}
  {462, 563}

\bibitem[\protect\citeauthoryear{{O{\~n}orbe}, {Boylan-Kolchin}, {Bullock},
  {Hopkins}, {Kere{\v s}}, {Faucher-Gigu{\`e}re}, {Quataert}  \&
  {Murray}}{{O{\~n}orbe} et~al.}{2015}]{Orbe2015}
{O{\~n}orbe} J.,  {Boylan-Kolchin} M.,  {Bullock} J.~S.,  {Hopkins} P.~F.,
  {Kere{\v s}} D.,  {Faucher-Gigu{\`e}re} C.-A.,  {Quataert} E.,   {Murray} N.,
   2015, \mn@doi [\mnras] {10.1093/mnras/stv2072}, \href
  {http://adsabs.harvard.edu/abs/2015MNRAS.454.2092O} {454, 2092}

\bibitem[\protect\citeauthoryear{{Papastergis} \& {Shankar}}{{Papastergis} \&
  {Shankar}}{2015}]{Papastergis2015}
{Papastergis} E.,  {Shankar} F.,  2015, preprint, \href
  {http://adsabs.harvard.edu/abs/2015arXiv151108741P} {} (\mn@eprint {arXiv}
  {1511.08741})

\bibitem[\protect\citeauthoryear{{Pawlowski}, {Famaey}, {Merritt}  \&
  {Kroupa}}{{Pawlowski} et~al.}{2015}]{Pawlowski2015}
{Pawlowski} M.~S.,  {Famaey} B.,  {Merritt} D.,   {Kroupa} P.,  2015, \mn@doi
  [\apj] {10.1088/0004-637X/815/1/19}, \href
  {http://adsabs.harvard.edu/abs/2015ApJ...815...19P} {815, 19}

\bibitem[\protect\citeauthoryear{{Perryman} et~al.,}{{Perryman}
  et~al.}{2001}]{Gaia}
{Perryman} M.~A.~C.,  et~al., 2001, \mn@doi [\aap]
  {10.1051/0004-6361:20010085}, \href
  {http://adsabs.harvard.edu/abs/2001A%26A...369..339P} {369, 339}

\bibitem[\protect\citeauthoryear{{Planck Collaboration} et~al.,}{{Planck
  Collaboration} et~al.}{2014}]{PlanckParams2013}
{Planck Collaboration} et~al., 2014, \mn@doi [\aap]
  {10.1051/0004-6361/201321591}, \href
  {http://adsabs.harvard.edu/abs/2014A%26A...571A..16P} {571, A16}

\bibitem[\protect\citeauthoryear{{Planck Collaboration} et~al.,}{{Planck
  Collaboration} et~al.}{2015}]{PlanckP2015}
{Planck Collaboration} et~al., 2015, preprint, \href
  {http://adsabs.harvard.edu/abs/2015arXiv150201589P} {} (\mn@eprint {arXiv}
  {1502.01589})

\bibitem[\protect\citeauthoryear{{Pontzen} \& {Governato}}{{Pontzen} \&
  {Governato}}{2014}]{Pontzen2014}
{Pontzen} A.,  {Governato} F.,  2014, \mn@doi [\nat] {10.1038/nature12953},
  \href {http://adsabs.harvard.edu/abs/2014Natur.506..171P} {506, 171}

\bibitem[\protect\citeauthoryear{{Sarkar}, {Mondal}, {Das}, {Sethi},
  {Bharadwaj}  \& {Marsh}}{{Sarkar} et~al.}{2015}]{Sarkar2015}
{Sarkar} A.,  {Mondal} R.,  {Das} S.,  {Sethi} S.~K.,  {Bharadwaj} S.,
  {Marsh} D.~J.~E.,  2015, preprint, \href
  {http://adsabs.harvard.edu/abs/2015arXiv151203325S} {} (\mn@eprint {arXiv}
  {1512.03325})

\bibitem[\protect\citeauthoryear{{Schive}, {Chiueh}  \& {Broadhurst}}{{Schive}
  et~al.}{2014a}]{Schive1}
{Schive} H.-Y.,  {Chiueh} T.,   {Broadhurst} T.,  2014a, \mn@doi [Nature
  Physics] {10.1038/nphys2996}, \href
  {http://adsabs.harvard.edu/abs/2014NatPh..10..496S} {10, 496}

\bibitem[\protect\citeauthoryear{{Schive}, {Liao}, {Woo}, {Wong}, {Chiueh},
  {Broadhurst}  \& {Hwang}}{{Schive} et~al.}{2014b}]{Schive2}
{Schive} H.-Y.,  {Liao} M.-H.,  {Woo} T.-P.,  {Wong} S.-K.,  {Chiueh} T.,
  {Broadhurst} T.,   {Hwang} W.-Y.~P.,  2014b, \mn@doi [Physical Review
  Letters] {10.1103/PhysRevLett.113.261302}, \href
  {http://adsabs.harvard.edu/abs/2014PhRvL.113z1302S} {113, 261302}

\bibitem[\protect\citeauthoryear{{Schive}, {Chiueh}, {Broadhurst}  \&
  {Huang}}{{Schive} et~al.}{2016}]{Schive3}
{Schive} H.-Y.,  {Chiueh} T.,  {Broadhurst} T.,   {Huang} K.-W.,  2016, \mn@doi
  [\apj] {10.3847/0004-637X/818/1/89}, \href
  {http://adsabs.harvard.edu/abs/2016ApJ...818...89S} {818, 89}

\bibitem[\protect\citeauthoryear{{Sikivie} \& {Yang}}{{Sikivie} \&
  {Yang}}{2009}]{Sikivie2009}
{Sikivie} P.,  {Yang} Q.,  2009, \mn@doi [Physical Review Letters]
  {10.1103/PhysRevLett.103.111301}, \href
  {http://adsabs.harvard.edu/abs/2009PhRvL.103k1301S} {103, 111301}

\bibitem[\protect\citeauthoryear{{Spergel} et~al.,}{{Spergel}
  et~al.}{2003}]{Spergel2003}
{Spergel} D.~N.,  et~al., 2003, \mn@doi [\apjs] {10.1086/377226}, \href
  {http://adsabs.harvard.edu/abs/2003ApJS..148..175S} {148, 175}

\bibitem[\protect\citeauthoryear{{Spergel} et~al.,}{{Spergel}
  et~al.}{2013}]{WFIRST}
{Spergel} D.,  et~al., 2013, preprint, \href
  {http://adsabs.harvard.edu/abs/2013arXiv1305.5422S} {} (\mn@eprint {arXiv}
  {1305.5422})

\bibitem[\protect\citeauthoryear{{Su{\'a}rez}, {Robles}  \&
  {Matos}}{{Su{\'a}rez} et~al.}{2014}]{Suarez2013}
{Su{\'a}rez} A.,  {Robles} V.~H.,   {Matos} T.,  2014, \mn@doi [Astrophysics
  and Space Science Proceedings] {10.1007/978-3-319-02063-1_9}, \href
  {http://adsabs.harvard.edu/abs/2014ASSP...38..107S} {38, 107}

\bibitem[\protect\citeauthoryear{{Turner}}{{Turner}}{1983}]{Turner1983}
{Turner} M.~S.,  1983, \mn@doi [\prd] {10.1103/PhysRevD.28.1243}, \href
  {http://adsabs.harvard.edu/abs/1983PhRvD..28.1243T} {28, 1243}

\bibitem[\protect\citeauthoryear{{Weinberg}, {Bullock}, {Governato}, {Kuzio de
  Naray}  \& {Peter}}{{Weinberg} et~al.}{2013}]{Weinberg2013}
{Weinberg} D.~H.,  {Bullock} J.~S.,  {Governato} F.,  {Kuzio de Naray} R.,
  {Peter} A.~H.~G.,  2013, preprint, \href
  {http://adsabs.harvard.edu/abs/2013arXiv1306.0913W} {} (\mn@eprint {arXiv}
  {1306.0913})

\bibitem[\protect\citeauthoryear{{White} \& {Frenk}}{{White} \&
  {Frenk}}{1991}]{White91}
{White} S.~D.~M.,  {Frenk} C.~S.,  1991, \mn@doi [\apj] {10.1086/170483}, \href
  {http://adsabs.harvard.edu/abs/1991ApJ...379...52W} {379, 52}

\bibitem[\protect\citeauthoryear{{White} \& {Rees}}{{White} \&
  {Rees}}{1978}]{White78}
{White} S.~D.~M.,  {Rees} M.~J.,  1978, \mnras, \href
  {http://adsabs.harvard.edu/abs/1978MNRAS.183..341W} {183, 341}

\bibitem[\protect\citeauthoryear{{Widrow} \& {Kaiser}}{{Widrow} \&
  {Kaiser}}{1993}]{Widrow93}
{Widrow} L.~M.,  {Kaiser} N.,  1993, \mn@doi [\apjl] {10.1086/187073}, \href
  {http://adsabs.harvard.edu/abs/1993ApJ...416L..71W} {416, L71}

\bibitem[\protect\citeauthoryear{{Willman} \& {Strader}}{{Willman} \&
  {Strader}}{2012}]{Willman2012}
{Willman} B.,  {Strader} J.,  2012, \mn@doi [\aj] {10.1088/0004-6256/144/3/76},
  \href {http://adsabs.harvard.edu/abs/2012AJ....144...76W} {144, 76}

\bibitem[\protect\citeauthoryear{{Wolf}, {Martinez}, {Bullock}, {Kaplinghat},
  {Geha}, {Mu{\~n}oz}, {Simon}  \& {Avedo}}{{Wolf} et~al.}{2010}]{Wolf2010}
{Wolf} J.,  {Martinez} G.~D.,  {Bullock} J.~S.,  {Kaplinghat} M.,  {Geha} M.,
  {Mu{\~n}oz} R.~R.,  {Simon} J.~D.,   {Avedo} F.~F.,  2010, \mn@doi [\mnras]
  {10.1111/j.1365-2966.2010.16753.x}, \href
  {http://adsabs.harvard.edu/abs/2010MNRAS.406.1220W} {406, 1220}

\bibitem[\protect\citeauthoryear{{Woo} \& {Chiueh}}{{Woo} \&
  {Chiueh}}{2009}]{Woo2009}
{Woo} T.-P.,  {Chiueh} T.,  2009, \mn@doi [\apj] {10.1088/0004-637X/697/1/850},
  \href {http://adsabs.harvard.edu/abs/2009ApJ...697..850W} {697, 850}

\makeatother
\end{thebibliography}

\end{document}